\documentclass[twocolumn]{aa}
\usepackage{graphicx}
\usepackage{txfonts}
\begin{document}
   \title{Mid-IR emission of galaxies in the Virgo cluster {\bf and in the Coma supercluster.IV. 
   The nature of the dust heating sources}}

   \author{A.Boselli\inst{1}, J.Lequeux\inst{2} and G.Gavazzi\inst{3}}

   \offprints{A.Boselli}

   \institute
             {Laboratoire d'Astrophysique de Marseille, BP8, Traverse du Siphon - F-13376 Marseille Cedex 12, France.\\
             \email{Alessandro.Boselli@oamp.fr}
	\and
	     Observatoire de Paris, LERMA, 61 Av. de l'Observatoire, 75014 Paris, France.\\
	     \email{James.Lequeux@obspm.fr}
	\and
	     Universit\'{a} degli Studi di Milano-Bicocca, P.zza della Scienza 3, 20126 Milano, Italy.\\
              \email{Giuseppe.Gavazzi@mib.infn.it}
       }

   \date{Received ......| Accepted ........}

   \abstract
{We study the relationship between the mid-IR (5-18 $\mu$m) emission of late-type galaxies and various other star 
formation tracers in order to investigate the nature of the dust heating sources in this spectral domain. 
The analysis is carried out using a sample of 123 normal, late-type, nearby galaxies with available data at several frequencies.
The mid-IR luminosity (normalized to the H-band luminosity) correlates better with the 
far-IR luminosity than with more direct tracers of the young 
stellar population such as the H$\alpha$ and the UV luminosity. The comparison of resolved images reveals a remarkable 
similarity in the H$\alpha$ and mid-IR morphologies, with prominent HII regions at both frequencies. 
The mid-IR images, however, show
in addition a diffuse emission not associated with HII regions nor with the diffuse H$\alpha$ emission. This evidence
indicates that the stellar population responsible for the heating of dust emitting in the mid-IR is similar to that
heating big grains emitting in the far-IR, including relatively evolved stars responsible for the non-ionizing radiation.
The scatter in the mid-IR vs. H$\alpha$, UV and far-IR luminosity relation is mostly due to metallicity effects, with metal-poor 
objects having a lower mid-IR emission per unit star formation rate than metal-rich galaxies.
Our analysis indicates that the mid-IR luminosity is not an optimal  
star formation tracer in normal, late-type galaxies.
   }

\titlerunning{mid-IR emission of galaxies in the Virgo cluster}
\authorrunning{A.Boselli et al.}

   \maketitle
%

\section{Introduction}

The atomic gas, a major constituent of proto-galaxies, collapsed into molecular clouds to form 
stars at a rate variable with time. At least a fraction of elliptical galaxies 
was formed during a violent, rapid ($\leq$ 10$^8$ yr) collapse 
of the primordial gas, efficiently transformed into stars, probably induced by merging events, frequent
in the dense early Universe. In spiral discs the rotation prevented a rapid collapse of the 
primordial gas, which formed a reservoir able to sustain the star formation up to the present epoch. 
The current star formation activity of nearby spiral galaxies seems primarily 
governed by the gas surface density, possibly modulated by the galactic differential rotation or by 
the stellar surface density (Kennicutt 1998a; Boissier et al. 2003). The star formation history of
these objects should thus have been monotonic, governed by the physical
properties of the original protogalaxies such as their angular momentum
(Sandage 1986) or total mass (Boselli et al. 2001).  \\
The study of the evolution of galaxies requires some
reconstruction of their star formation activity from their birth to the present.
In the nearby Universe the present day star formation activity of normal, late-type galaxies can be
easily determined by combining H$\alpha$ or UV data with population synthesis models
(Kennicutt 1998a; Boselli et al. 2001). The major limitation in this technique is 
the determination of the dust extinction. Low-resolution integrated spectroscopy 
in the visible domain, now available for large galaxy samples 
(Kennicutt 1992; Jansen et al. 2000; Gavazzi et al. 2004), as well as far-IR data from the IRAS
survey, however, allow us to accurately quantify
the amount of extinction using the Balmer decrement or the far-IR to UV flux ratio (Buat et al. 2002).\\
Quantifying the star formation activity of galaxies at higher redshift is a more complex task.
For $z$$\geq$0.5 the H$\alpha$ line shifts outside the optical domain, entering the 
less accessible near-IR part of the spectrum, where the atmospheric opacity and the sky instability make
the determination of the H$\alpha$ line intensity and of the Balmer decrement more difficult.
Deep optical observations provide the UV rest-frame photometry, but the determination of the dust extinction,
necessary to transform the UV flux into a quantitative star formation rate, can be accomplished only when 
far-IR data are known. These will soon be available for small patches of the sky from the 
SWIRE-Spitzer survey ($\sim$ 70 deg$^2$).
Since at redshifts higher than $z$ $\sim$ 1 most of the star formation should manifest 
itself by a major starburst inside the galaxy 
(Steidel et al. 1999), where extinction is particularly high, Balmer line and UV luminosities 
are not reliable star formation indicators. As shown by Kennicutt (1998a,b), for these
highly extincted starburst galaxies the far-IR emission becomes the best star formation indicator.
Far-IR data presently available from IRAS, ISO, Spitzer and shortly from ASTRO-F are however limited by 
the poor spatial resolution, of the order of 1.5-0.5 $\arcmin$, making confusion a major limitation.\\
Mid-IR data have been proposed as a promising alternative star formation tracer, particularly suitable
for highly extincted environments because less affected by obscuration than other indicators at shorter wavelength 
(Roussel et al. 2001; F\"orster Schreiber et al. 2004).
Compared to far-IR, the mid-IR data have the advantage of a superior spatial resolution ($\sim$ a
factor of 10 depending on wavelength). Furthermore, the mid-IR emission of late-type galaxies is expected to
be more tightly related to the young stellar population than the far-IR one (D\'esert et al. 1990) 
thus representing a potentially more direct star formation tracer. In quiescent spiral galaxies, indeed, 
the far-IR emission is known to depend also on the old stellar population via circumstellar and/or photospheric
emission or diffuse cirrus emission (Sauvage \& Thuan 1994).\\
The one to one relationship between mid-IR emission and recent star formation, 
was however questioned by Boselli et al. (1997a; 1998) who found a lack of 
correlation between the mid-IR emission at 6.75 and 15 $\mu$m and various star formation tracers, and concluded 
that the carriers of the UIBs responsible for the emission at these wavelengths might be destroyed in high UV 
radiation fields, as first suggested by Boulanger et al. (1988) and Helou et al. (1991).\\ 
To understand the reason for this evident discrepancy, it is important to investigate in depth, both empirically and theoretically,
the relationship between the mid-IR emission and the star formation rate.
The emission of the ISM in the mid-IR is due to very small grains (VSG; $10 \AA \la a \la 200 \AA$) responsible for the
continuum, and to smaller particles, not in thermal equilibrium with the radiation, responsible for the Unidentified 
Infrared Bands (UIBs).The mid-IR emission can also be due to  
H$_2$ rotational lines, to fine-structure lines of various metals 
and H recombination lines produced in HII and photodissociation regions (Sturm et al. 2000)
(the latter contribute only to a small fraction of the flux in the ISOCAM bands centered at 
6.75 and 15 $\mu$m and can be neglected). 
The small grains are stochastically heated to very high temperatures by the absorption 
of individual photons. 
UIBs are supposedly associated to polycyclic aromatic hydrocarbon molecules (PAHs; L\'eger \& Puget 1984; D\'esert et al. 1990) 
or to hydrogenated amorphous carbon grains (Duley \& Williams 1981; 1988).  
The study of the mid-IR spectral energy distribution (SED) and of its relationships with the UV radiation field
in various extragalactic (IR bright galaxies, active galactic nuclei (AGN), normal late-type galaxies, blue compact dwarf
galaxies (BCDs), ellipticals)
and galactic environments (HII regions, diffuse medium, reflection nebulae,...) (see for a review Genzel \& Cesarsky 2000)
was recently boosted by the ISO and COBE missions. Spectroscopic studies 
showed that UIBs dominate the spectrum in the range 6-13 $\mu$m of normal (Lu et al. 2003)
and starburst (F\"orster Schreiber et al. 2003) galaxies, as well as in Galactic HII and photodissociation regions
(Roelfsema et al. 1996; Cesarsky et al. 1996; Verstraete et al. 1996, 2001), planetary and reflection nebulae 
(Beintema et al. 1996; Boulanger et al. 1996; Uchida et al. 1998) and in the diffuse medium associated to dust cirrus
(Mattila et al. 1996; Lemke et al. 1998; Chan et al. 2001). VSGs dominate the emission at longer wavelengths but also
in the 6-13 $\mu$m regime whenever the UV radiation field is extremely high and/or the metallicity is low, as in BCDs
and Wolf-Rayet galaxies (Madden 2000; Galliano et al. 2003; Crowther et al. 1999), AGN (Roche et al. 1991; 
Dudley 1999; Laurent et al. 2000), or close to prominent HII regions in low metallicity environments, 
such as N66 in the SMC (Contursi et al. 2000).\\
Further compelling evidence for the lack of direct relation between mid-IR dust emission and star formation
comes from detailed studies of the physical properties of the ISM in the Milky Way and in some nearby galaxies.
These studies showed that the presence of UIBs in the mid-IR spectrum is not necessarily associated to
strong UV radiation, indicating that the carriers of the UIBs can be effectively excited also
by less energetic photons produced by relatively evolved stars (Uchida et al. 1998; Pagani et al. 1999).\\
The complex behavior of the different components of the ISM observed in our Galaxy or in a few well studied
nearby objects has made it possible to realize how difficult it is to interpret the global emission of 
unresolved galaxies observed with broad band filters, thus without detailed spectral information, when all 
the various components of the ISM (HII regions, cirrus, photo-dissociation regions (PDRs), diffuse medium)
are mixed together and the properties of the interstellar radiation field change from object to object.\\
With this aim in mind, e.g. to better quantify if the mid-IR emission is a reliable star formation  
tracer in late-type galaxies, we propose in this work a re-discussion, based on 
a multifrequency statistical approach, of the nature of the dust heating sources
responsible for the  mid-IR emission in normal galaxies in conjunction with the properties of 
the UV radiation field and with the metallicity of the ISM. 
The major improvement with respect to our previous analysis (Boselli et al. 1997a; 1998) on this issue
lies in the use of a better and larger mid-IR data-set homogeneously reduced using the latest ISOCAM data reduction pipeline
and the availability of better imaging and spectroscopic material. \\
This work is the natural continuation of earlier statistical analyses
(Boselli et al. 1997a; 1998) based on
a large sample of late-type galaxies in the Virgo cluster that were
observed by the ISOCAM consortium in guaranteed time and on several objects in the 
Coma supercluster region (Contursi et al. 2001).  
A revised version of this data set was presented in Boselli et al. (2003a) and was discussed by Boselli et al. (2003b) 
who reconstructed the UV-to-centimetric spectral energy distribution (SED) of normal, late-type galaxies. 
In summary, the analysis completed so far indicated that the system mass rather than 
the morphological type regulates the mid-IR emission (Boselli et al. 1998) and that
in late-type normal galaxies 
the contribution of the cold stellar component to the total mid-IR emission is relevant, becoming
dominant in quiescent, early-type spirals.

\section{The sample}

The study of the mid-IR properties of galaxies is mostly limited
by the lack of data for homogeneous, complete and large samples 
of galaxies in the nearby Universe. The present work is based on
the analysis of the sample of galaxies in the Virgo cluster 
and in the Coma/A1367 supercluster region by Boselli et al. (2003a).
This sample, which includes 145 objects with
homogeneously reduced ISOCAM 6.75 and 15 $\mu$m imaging data of which 123 are late-type objects, 
is not complete. However, as extensively discussed in Boselli et al. 
(2003a, 2003b), it includes an optically selected, volume limited, 
complete sample of 100 late-type galaxies in the Virgo cluster.  
The following analysis will be limited to the 123 late-type galaxies of type Sa-Im-BCD 
in the luminosity range -21 $\leq$ $M_B$ $\leq$ -13.

\section{The data}

Mid-IR 6.75 (LW2) and 15 $\mu$m (LW3) ISOCAM imaging data have been taken from 
Boselli et al. (2003a). To avoid large uncertainties in the mid-IR data, 
we considered as detections only those with quality 1 and 2 
in Table 2 of Boselli et al. (2003a). As discussed in Boselli et
al. (2003a), it is difficult to estimate the effective uncertainty in the
mid-IR fluxes; here we assume a conservative 30\%.\\
Mid-IR luminosities at 6.75 and 15 $\mu$m (in solar units) 
are estimated as in Boselli et al. (1998): 

\begin{equation}
{L_{6.75 \mu m} = 4 \pi D^2 F_{6.75 \mu m} 2.724 \times 10^2 L_{\odot}}
\end{equation}

\noindent
and

\begin{equation}
{L_{15 \mu m} = 4 \pi D^2 F_{15 \mu m} 1.184 \times 10^2 L_{\odot}}
\end{equation}

\noindent
where $D$ is the distance in Mpc, and $F_{6.75 \mu m}$ and $F_{15 \mu m}$
are the mid-IR flux densities in mJy (for a bandwidth of $\delta_{6.75 \mu m}$=11.57 $\times$ 10$^{12}$ Hz
and $\delta_{15 \mu m}$=5.04 $\times$ 10$^{12}$ Hz respectively).\\
Three different tracers are used to estimate the star formation activity of the target galaxies: 
the H$\alpha$, the UV and the far-IR luminosity.
H$\alpha$+[NII] imaging and/or aperture photometry data, available
for 96 galaxies, are taken from
Boselli et al. (2002a; 2003b), Boselli \& Gavazzi (2002), Gavazzi et al. (1991;
1998; 2002b), Iglesias-Paramo et a. (2002), Kennicutt \& Kent (1983), Kennicutt et al. (1984) and 
references therein \footnote{Fluxes from 
Kennicutt \& Kent (1983) data are increased
by 16 \% as specified in Kennicutt et al. (1994) (also Kennicutt private
communication) to account for a telluric line contamination in the off
band filter}.\\
Spectroscopic data, necessary to correct H$\alpha$ imaging data for [NII]
contamination and dust extinction taken from the survey of Gavazzi et al. (2004), are
available for 91 objects.
Long slit, drift-scan mode spectra were obtained by drifting the slit over the
whole galaxy disc, as in Kennicutt (1992), thus providing values representative of
the whole object. These are intermediate ($\lambda/\Delta\lambda \sim$ 1000) 
resolution spectra in the range ($3600 - 7200$ \AA).  \\
H$\alpha$ imaging and aperture photometry data
were corrected for dust extinction using the Balmer decrement given
in Gavazzi et al. (2004), determined for 78 out of the 91 galaxies using available
spectroscopic data. This correction is quite accurate since it takes into
account the contribution of the underlying Balmer absorption to H$\beta$.
For those objects without integrated spectroscopy or measured Balmer decrement (20\%) we applied
the corrections proposed by Boselli et al. (2002b)
for the [NII] contamination ($Log([NII]/H\alpha$)=0.35 $\times$ $Log L_H$ -3.85), 
and Boselli et al. (2001) for extinction:
$A({\rm H}\alpha)$=1.1 magnitudes for Sa-Scd and Pec galaxies, and
$A({\rm H}\alpha)$=0.6 mag for the other objects. Consistently with Iglesias-Paramo et al. (2004)
we also assume that the fraction of ionized photons absorbed by dust is $\sim$ 40 \%, leading to
a fraction of ionizing photons absorbed by the gas $f$=0.6.
The mean error on the H$\alpha$ luminosity is $\sim$ 15\%.\\
The spectroscopic survey of Gavazzi et al. (2004) was also used to 
estimate metallicities as  $12~+~Log(O/H)$, available for 68 objects. Metallicities have been 
estimated by averaging various calibrations, namely 
van Zee et al. (1998), Kewley \& Dopita (2002), Kobulnicky et al. (1999), McGaugh (1991), Dutil \& Roy (1999).\\
The UV data are taken from observations with FAUST (Deharveng et al. 1994, at 1650 \AA) and FOCA 
(Donas et al. 1991; 1995, at 2000 \AA).
As in Boselli et al. (2003b) to be consistent with the FOCA data we transform FAUST UV magnitudes taken 
at 1650 \AA~ to 2000 \AA~ assuming a 
constant color index UV(2000)=UV(1650)+0.2 mag. UV magnitudes used in the following analysis
thus refer to a wavelength of 2000 \AA. 
These are total magnitudes, determined by integrating the UV emission up to
the weakest detectable isophote. The average estimated error in the UV magnitude is 
0.3 mag, but it ranges from 0.2 mag for bright galaxies to 0.5 
mag for weak sources observed in frames with larger than average calibration 
uncertainties.
UV data are corrected for dust attenuation as described in Boselli et
al. (2003b): 82 \% of the 62 galaxies detected at 60 and 100 $\mu$m by IRAS or
ISOPHOT have UV data: for these objects the UV extinction is determined from the FIR to UV
flux ratio using the calibration given in Boselli et al. (2003b). 
Objects undetected at 60 or 100 $\mu$m were corrected using the average 
UV extinction determined for galaxies of similar morphological type:
$A(UV)$ = 1.28; 0.85; 0.68 mag
for Sa-Sbc; Sc-Scd; Sd-Im-BCD galaxies respectively.\\
Far-IR IRAS fluxes at 60 and 100 $\mu$m have been taken from different 
sources. Alternative ISOPHOT far-IR values at 60 and 100 $\mu$m have been 
taken from Tuffs et al. (2002), bringing to 62 the number of objects with
available far-IR data. The comparison of ISO and IRAS data for 
the sample galaxies detected in both surveys reveals a systematic 
difference of ISO/IRAS=0.95 and 0.82 at 60 and 100 $\mu$m respectively 
(Tuffs et al. 2002). To be consistent with IRAS, ISOPHOT data have
been multiplied by these factors. Far-IR luminosities are determined from the 60 and
100 $\mu$m flux densities as in Boselli et al. (2003b). The mean uncertainty in
the far-IR data is 15 \%.\\
An accurate determination of the H$\alpha$, UV and far-IR uncertainties determined
with a similar set of data is given in Iglesias-Paramo et al. (2004).\\
To compare galaxies of different sizes we remove the well known luminosity-luminosity
or luminosity-mass scaling relations by normalizing luminosities in all bands
to the H band luminosity, which is a good tracer of the total dynamical mass
of late-type galaxies (Gavazzi et al. 1996a). 
Near-IR data have been taken from our H and K' band 
surveys of the Virgo cluster and of the Coma/A1367 supercluster 
(Boselli et al. 1997b, 2000; Gavazzi et al 1996b,c; 2000a,b).
Total H band luminosities have been determined using the relation:
$Log L(H)$=11.36-0.4 $\times$ $H_T$ $+$ 2 $Log D$, where $H_T$ is 
the total H band magnitude and $D$ is the distance in Mpc. 
Total extrapolated near-IR magnitudes have been determined as 
described in Gavazzi et al. (2000a), with a mean uncertainty of $\sim$ 10\%. 
K' magnitudes have been transformed into
H magnitudes adopting a constant H-K'=0.25 (independent of type; see
Gavazzi et al. 2000a) when the color index is not available. This is an 
additional source of uncertainty.\\
We recall that the quoted uncertaintes are mean values: the uncertainties in the 
spectro-photometric quantities of individual galaxies are larger and represent a 
source of dispersion in the relationships analysed throughout this work.\\
As shown by Boselli et al. (1998, 2003b), the emission of late-type galaxies
in the mid-IR is partly contaminated by stellar emission. At 6.75 $\mu$m 
the emission of the photosphere of the cold stellar population is dominant 
in Sa galaxies, where it reaches on average $\sim$ 80 \% of the total emission, 
while its contribution decreases in later types: in Sc it is on average $\sim$ 20 \%, 
and $\sim$ 50\% in BCDs.\\
As proposed by Boselli et al. (2003b), the stellar contribution at 6.75 $\mu$m
can be estimated from the SED determined by fitting the UV to near-IR spectrophotometric
data with the updated version of Bruzual \& Charlot population synthesis models (Bruzual \& Charlot 1993; GISSEL 2001). 
Owing to the available multifrequency dataset, this can be achieved for most (75) of the galaxies. 
For the remaining (48) objects
the pure dust emission has been estimated by subtracting from the 
6.75 $\mu$m emission the average stellar contribution for a given
morphological type, as determined from the template SED given in Boselli 
et al. (2003b). 
We remind the reader that the evolutionary synthesis models used in this work
to estimate the integrated mid-IR emission of late-type galaxies 
rely on either empirical stellar libraries and model atmospheres that 
are poorly known for cool stars, especially for evolved post-main-sequence stars
dominating the galaxy emission at this wavelength. Another source of error
of the fitting procedure is associated to the
uncertainty in the star formation history and metallicity assumed for the model. \\
The accuracy of the morphological classification is excellent for the 
Virgo galaxies (Binggeli et al. 1985; 1993).
Because of the larger distances, the morphology of galaxies belonging 
to the other surveyed regions suffers from an uncertainty 
of about 1.5 Hubble type bins.
We assume a distance of 17 Mpc for the members (and possible members) 
of Virgo cluster A, 22 Mpc for Virgo cluster B, and 32 Mpc for
objects in the M and W clouds (see Gavazzi et al. 1999).
Members of the clusters Coma and A1367 are assumed to be at 
distances of 86.6 and 92 Mpc respectively.  
Isolated galaxies in the Coma supercluster are assumed to be
at their redshift distance adopting $H_{0}$ = 75 km s$^{-1}$ Mpc$^{-1}$.
Most of the data used in this analysis are
available in the GOLDMine database (http//goldmine.mib.infn.it; Gavazzi et al. 2003).

\begin{table}
\caption{The number of objects with available data in various bands and the completeness
with respect to the 123 late-type galaxies considered in this work. }
\label{Tab1}
\[
\begin{array}{lc}
\hline
\noalign{\smallskip}
{\rm data} 		& {\rm n.~of~objects ~(\%)}\\
\hline
{\rm H\alpha} 		& 96~(78\%)  \\
{\rm UV} 		& 57~(46\%)  \\
{\rm FIR~(detected)} 	& 62~(50\%)  \\
{\rm NIR} 		& 114~(93\%) \\
{\rm Spectroscopy} 	& 91~(74\%)  \\
{\rm Balmer~dec.} 	& 78~(63\%)  \\
{\rm Metallicity}	& 68~(55\%)  \\
{\rm SED~for~stellar~correction}&75~(61\%) \\
\noalign{\smallskip}
\hline
\end{array}
\]
\end{table}

\section{Analysis}

\subsection{Star formation tracers}

The multifrequency dataset outlined above is well suited for studying the nature and the relative 
contribution of the different heating sources of the dust emitting in the mid-IR.
Three different tracers of star formation, the H$\alpha$, UV and FIR luminosities, can be determined from the
available data.\\
The H$\alpha$ luminosity gives a direct measure of the global photoionization rate of the 
interstellar medium due to high-mass ($m$ $>$ 10 ${\rm M\odot}$), young 
($\le$ 10$^7$ years) O-B stars (Kennicutt 1983; Kennicutt et al. 1994; Boselli et al. 2001).
If the SFR is constant over a time scale of some 
10$^7$ years (the life time of the ionizing stars), H$\alpha$ luminosity 
can be converted into SFR (in M$\odot$ yr$^{-1}$) assuming an initial mass function and
using population synthesis models, as extensively described in Boselli et al. (2001).\\
The UV emission of galaxies at 2000 \AA~ is dominated by 
less young ($\sim$ 10$^8$ years) and massive (2$<$ $m$ $<$ 5 ${\rm M\odot}$)
A stars (Lequeux 1988). If the SFR is stationary on time scales $\geq$ 3 10$^8$ years
\footnote{This assumption is reasonable for normal galaxies like those analyzed in this
work (Boselli et al. 2001; Gavazzi et al. 2002a).},
as for the H$\alpha$, UV luminosities can be converted into SFR using population synthesis
models.
These two tracers are thus independent. The major limitation of both indicators, however, 
is dust attenuation.
Although both tracers are accurately corrected using totally independent recipes
(the extinction in H$\alpha$ is determined from the Balmer decrement and in the 
UV from the far-IR to UV flux ratio), the SFR determined from these two indicators 
can suffer from similar biases and limitations.\\
The far-IR emission due to the dust heated by the stellar radiation field, although not directly
tracing the emission of the young stellar population, has the advantage of being extinction-free. 
As extensively discussed in the literature (Kennicutt 1998b; Sauvage \& Thuan 1994
and references therein),
the far-IR emission is an accurate star formation tracer only in strong starburst galaxies, where
most of the stellar radiation, particularly that emitted by the young, massive stars,
is absorbed by dust and re-radiated in the IR. In normal galaxies like those analyzed in this work, however,
the contribution of the dust heating by the radiation emitted by the older
stellar population can be significant, in particular in the most quiescent objects 
(Sauvage \&  Thuan 1994).\\
The correlations between the three tracers are shown in Fig. 1.
As previously mentioned, the strong relationship between the H$\alpha$, UV and far-IR
luminosities (left panels) is primarily a scaling effect 
(bigger galaxies have more of everything). For this reason the same relationships
are also given normalized to the near-IR luminosity in the right column.
Note that the dispersions in the UV vs. H$\alpha$ and FIR vs. UV relationships are significantly 
smaller than that in the FIR vs. H$\alpha$ relationships.
While the small dispersion in the UV vs. H$\alpha$ luminosity relationship is expected
because both quantities trace the young stellar population and suffer from similar extinction effects
\footnote{The dispersion in the UV vs. H$\alpha$ luminosity relationship
is probably partly due to uncertainties in the dust extinction correction and partly due
to recent star formation events (Iglesias-Paramo et al. 2004).},
that between the FIR and the UV luminosity might reflect the fact that
UV and FIR luminosities are not entirely independent entities, because
FIR data have been used to correct the UV luminosity for dust attenuation.

\begin{figure*}[]
\centering
\includegraphics[width=13.cm]{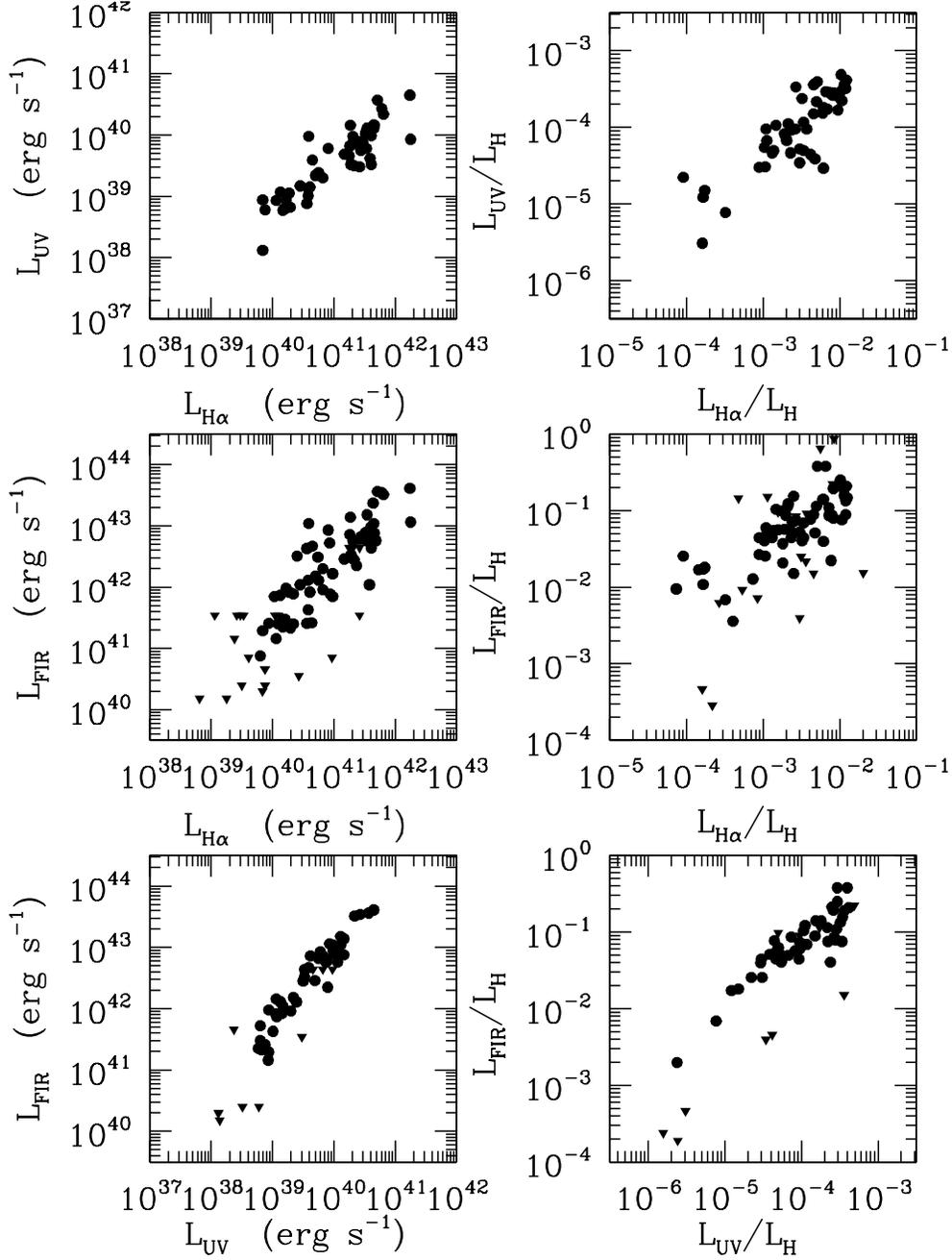}
\caption{The relationship between the H$\alpha$, the UV and the far-IR luminosities (left
panels) and normalized to the near-IR luminosity (right panel).
Triangles indicate upper limits to the far-IR luminosity.}
\label{lhauv}
\end{figure*}

\subsection{The origin of the mid-IR emission in late-type galaxies}

\subsubsection{Integrated values}

To study the origin of the mid-IR emission of normal, late-type galaxies we
compare in Fig. 2 the normalized 6.75 and 15 $\mu$m luminosity with that of the three 
star formation tracers previously analyzed. 

\begin{figure*}[]
\centering
\includegraphics[width=15.cm]{f2.epsi}
\caption{The relationship between the normalized (to the H band) mid-IR luminosities at 6.75 (upper) and 
15 $\mu$m (lower) and the H$\alpha$, the UV and the far-IR luminosity.
Filled symbols are for those objects whose mid-IR emission is dominated by dust 
($L_D+L_S/L_S(6.75 \mu m)$ $>$2), open symbols for galaxies whose mid-IR emission 
is mostly stellar ($L_D+L_S/L_S(6.75 \mu m)$ $\leq$ 2), which we consider the most doubtful data ($L_D$ stands for
dust luminosity, $L_S$ for stellar luminosity).
Triangles indicate upper limits to the mid-IR luminosity.}
\label{camnorm}
\end{figure*}
\noindent

Figure 2 shows a general correlation between the mid-IR emission
(in both bands) and the three star formation tracers. The dispersion 
is however small only for the correlation with the far-IR luminosity. 
If we limit the comparison to
objects whose mid-IR emission is dominated by dust (filled symbols in Fig. 2), avoiding
a large unwanted stellar contamination, the
dynamical range in the mid-IR luminosity is significantly reduced, the
correlation with the H$\alpha$ strongly weakens, and that with the UV luminosity almost vanishes.\\
We thus need to correct the
mid-IR emission for the stellar contribution. This correction can be achieved at 6.75 $\mu$m, 
since the stellar emission can be estimated from the Bruzual \& Charlot
models fitted to the stellar SEDs (Boselli et al. 2003b), see Sect. 3.
Throughout Figs. 2, 3, 6, 9, 10 
we keep separate the galaxies with mid-IR emission dominated by dust ($L_D+L_S/L_S(6.75 \mu m)$ $>$ 2) 
(filled symbols), requiring minor corrections to remove the stellar contribution, 
from objects whose emission is almost entirely stellar ($L_D+L_S/L_S(6.75 \mu m)$ $\leq$ 2),
so that their mid-IR dust emission is highly uncertain (empty symbols); in the above
$L_D$ stands for pure dust and $L_S$ for stellar luminosity. 
A similar correction cannot be applied at 15 $\mu$m since the Bruzual \& Charlot
models are limited to $\lambda$ $\leq$ 10 $\mu$m: the stellar contamination at wavelengths larger than
10 $\mu$m should however be minor.\\
The pure dust luminosity at 6.75 $\mu$m, $L_{D 6.75 \mu m}$ (normalized to the total H band luminosity), 
vs. the three star formation tracers is plotted in Fig. 3. There is still a general 
correlation between the mid-IR dust emission and the far-IR emissivity, although weaker than when uncorrected data are used, but the 
correlations with UV and H$\alpha$, if any, have become extremely weak.

\begin{figure*}[]
\centering
\includegraphics[width=15.cm]{f3.epsi}
\caption{The relationship between the normalized (to the H band) 
mid-IR luminosity at 6.75 corrected for the stellar contribution and the H$\alpha$, the UV and the far-IR luminosity.
Symbols as in Fig. 2.
}
\label{camdsfr}
\end{figure*}
\noindent

These trends however rely on the most doubtful 6.75 $\mu$m measurements (empty
symbols in Fig. 3).\\
It is remarkable that the dispersion in the $L(6.75 \mu m)/L(H)$
vs. $L(H\alpha)/L(H)$ relation is larger than in the $L(FIR)/L(H)$ vs. 
$L(H\alpha)/L(H)$ relation (Fig. 1). This suggests that the dispersion in the
$L(6.75 \mu m)/L(H)$ vs. $L(H\alpha)/L(H)$ relation is independent of the extinction
because errors in the extinction correction would introduce
a similar scatter in both relations.\\


\subsubsection{Resolved galaxies}

We further investigate the origin of the mid-IR emission in late-type galaxies
by comparing the 2-D 6.75 (and 15 $\mu$m images) of galaxies resolved by ISOCAM with
the images tracing the distribution of the young stellar population. Because of the poor
ISOCAM spatial resolution (6 $\arcsec$ at 6.75 $\mu$m), the number of fully resolved  objects 
showing some structure (nucleus, spiral arms, compact regions)
is limited to a dozen galaxies. 
Images tracing the distribution of the young stars
refer at present to the H$\alpha$+[NII] images (available for
all the sample galaxies). UV images will shortly be available from GALEX, while the far-IR images from IRAS do not have 
enough resolution ($\sim$ 1.5 $\arcmin$) for a meaningful comparison.\\
\begin{figure}[]
\centering
\includegraphics[width=8.cm]{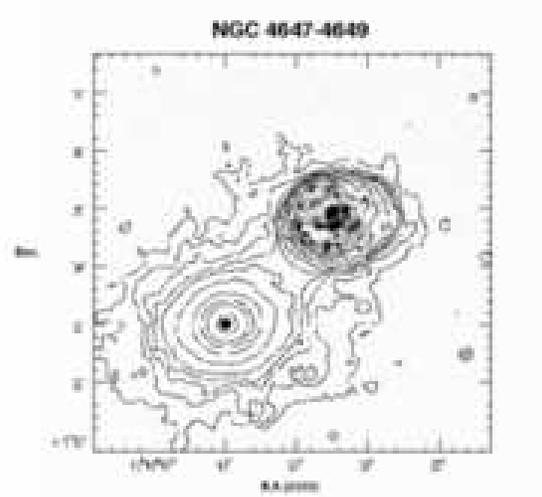}
\caption{The contour plot of the 6.75 $\mu$m emission (dust and stars) 
superimposed on the grey scale H$\alpha$+[NII] net image of the galaxies NGC 4647 and 4649.
Contours are given at 2,4,6,8,10,15,25,40,60,150,300 $\mu$Jy arcsec$^{-2}$. The grey scale is in arbitrary units.}
\label{ellittica}
\end{figure}
\noindent
Once again mid-IR images must be corrected 
for the stellar contamination. The 6.75 $\mu$m emission of some of the resolved galaxies of the sample of Boselli et al. (2003a)
is in fact totally stellar, as it is not associated to any recent star formation event. As an example the elliptical 
galaxy NGC 4649, companion of the spiral NGC 4647, has an extended emission in the mid-IR, but no recent 
star formation activity, as shown in Fig. 4.
As in Boselli et al. (2003a), we reconstruct the pure 6.75 $\mu$m emitting dust distribution
by assuming that the 2-D distribution of stars emitting at $\sim$ 7 $\mu$m is similar
to that at $\sim$ 2 $\mu$m. Thus we subtract from the 6.75 $\mu$m
images those in the K' band (appropriately scaled in flux using the fitted UV-to-near-IR SED
and smoothed to the same resolution \footnote{We thus make the reasonable assumption that, on large scales,
the effect due to an extinction gradient can be neglected.}). As previously remarked, 
this correction cannot be applied to the 15 $\mu$m images: at these wavelengths, however,
we can consider that the stellar contribution is only marginal.
The contour plots of the 6.75 $\mu$m dust emission (uncontaminated by stars) of the 
resolved galaxies are superimposed on the grey scale H$\alpha$+[NII]{\bf \footnote{As for the integrated values,
to trace star formation H$\alpha$+[NII] images should be corrected for dust extinction and [NII] contribution.
Although a radial variation of both entities is expected (Boissier et al. 2004), this should be minor, thus  
affecting the star forming morphology of these galaxies only marginally.}} net images in Fig. 5 \footnote{
The 15 $\mu$m images of the same galaxies, not given here, show morphologies similar to those at 6.75 $\mu$m.}. 
\noindent
Despite the difference in spatial resolution (6 $\arcsec$ for ISOCAM vs.
$\sim$ 1-2.5 $\arcsec$ for H$\alpha$), the ionizing flux and the mid-IR images show a
remarkably similar morphology. All bright spots resolved at 6.75 $\mu$m are clearly associated
with star forming HII regions. In addition, however, the ISOCAM images show a low-surface-brightness 
diffuse emission extending over the optical disc of the galaxies 
not associated with star forming HII regions nor with diffuse H$\alpha$ emission. 
Although radiative transfer processes cannot be totally excluded, in relatively quiescent 
galaxies such as the Milky Way (see next section) the disc interstellar radiation field outside 
HII regions is not necessarily produced by the youngest stellar population, but 
by stars showing a range of ages and masses (Uchida et al. 1998; Pagani et al. 1999). This suggests that
PAHs and VSG, dominating the emission of galaxies in the mid-IR, are mostly heated by the young stellar
population inside HII regions, while they are heated by all stellar populations all throughout their discs.

\section{Discussion}

The analysis carried out so far on both resolved and unresolved galaxies 
led us to the conclusion that the carriers of the UIB and associated continuum
responsible for the integrated mid-IR emission of normal, late-type galaxies 
are heated by both the UV ionizing and non-ionizing radiation emitted 
by the young and old stellar populations. This evidence indicates that the
dust emitting in the mid-IR is heated in the same way as the big grains emitting in the far-IR for which the contribution of 
visible and near-IR photons produced by low-mass old stars is well established.
This result agrees with what is found 
in the Milky Way and in M31, where UIBs have been detected in the diffuse ISM in regions 
with low UV radiation (Sellgren et al. 1990; Mattila et al. 1996; 
Uchida et al. 1998; Pagani et al. 1999; Li \& Draine 2002a).\\ 
It is difficult to know to what extent the contribution of the non-ionizing radiation 
differs in galaxies of different morphological type
and/or luminosity. The emission of dust at 6.75 $\mu$m is only weakly related to
the morphological type, increasing from Sa to Sd, Im and peculiar galaxies, but decreasing
again in BCDs, as shown in Fig. 6. This may be due to a lower abundance of dust in those
metal-poor galaxies. The relationship with the total mass of galaxies is even more complex: 
the $L_{D 6.75 \mu m}/L_H$ ratio increases with the H band
luminosity from $L_H$ = 10$^8$ to 10$^{10.5}$ L$_{H\odot}$ (filled dots in Fig. 6), but might 
decrease at higher luminosities. It seems therefore that the mid-IR pure dust emission is
more intense in massive, quiescent, metal-rich objects than in star forming, metal-poor 
dwarfs.\\

\begin{figure*}[]
\centering
 \includegraphics[width=7.5cm]{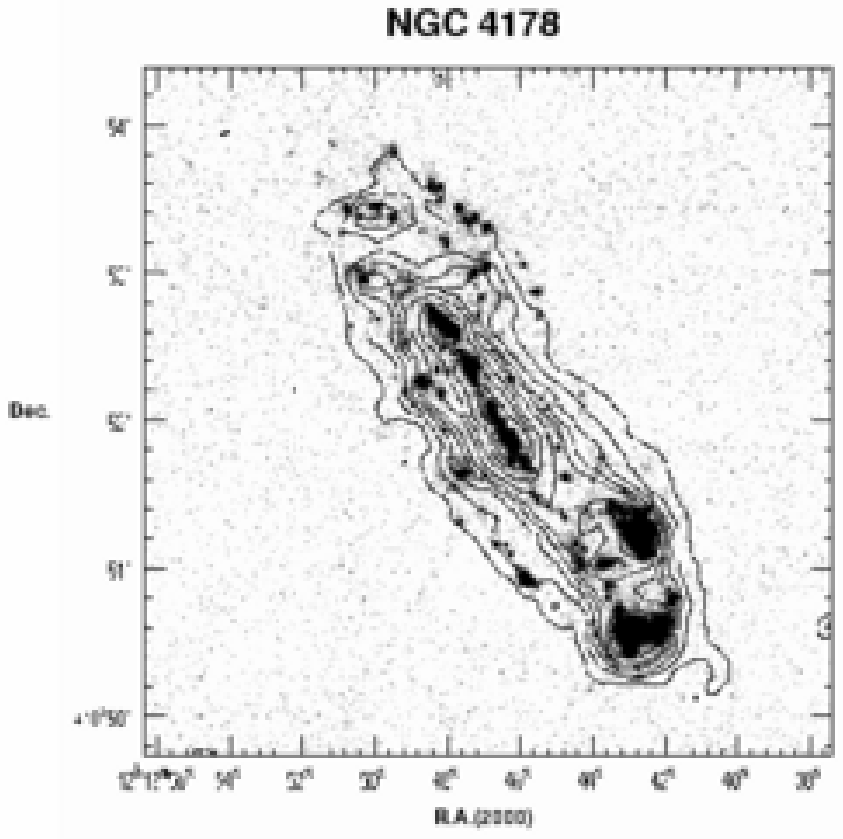}
 \includegraphics[width=7.5cm]{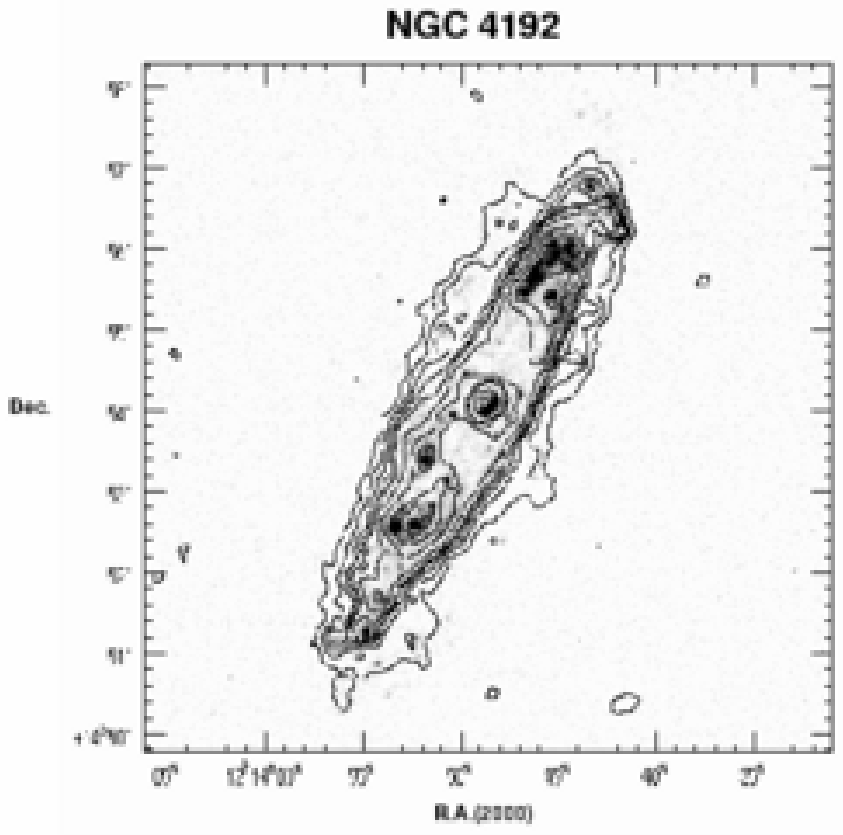}
 \includegraphics[width=7.5cm]{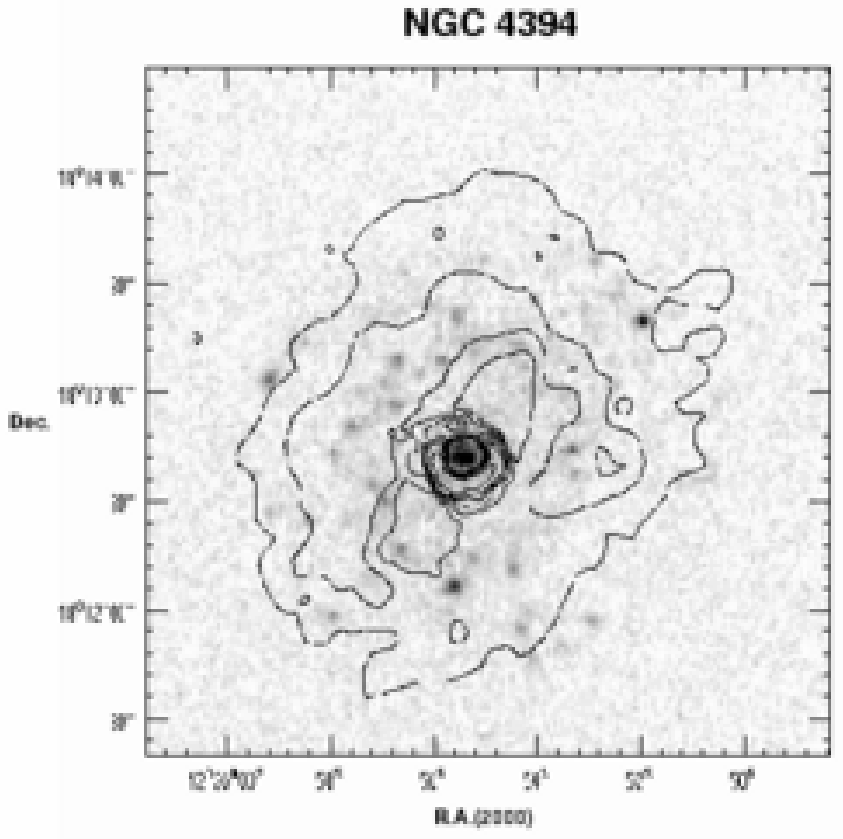}
 \includegraphics[width=7.5cm]{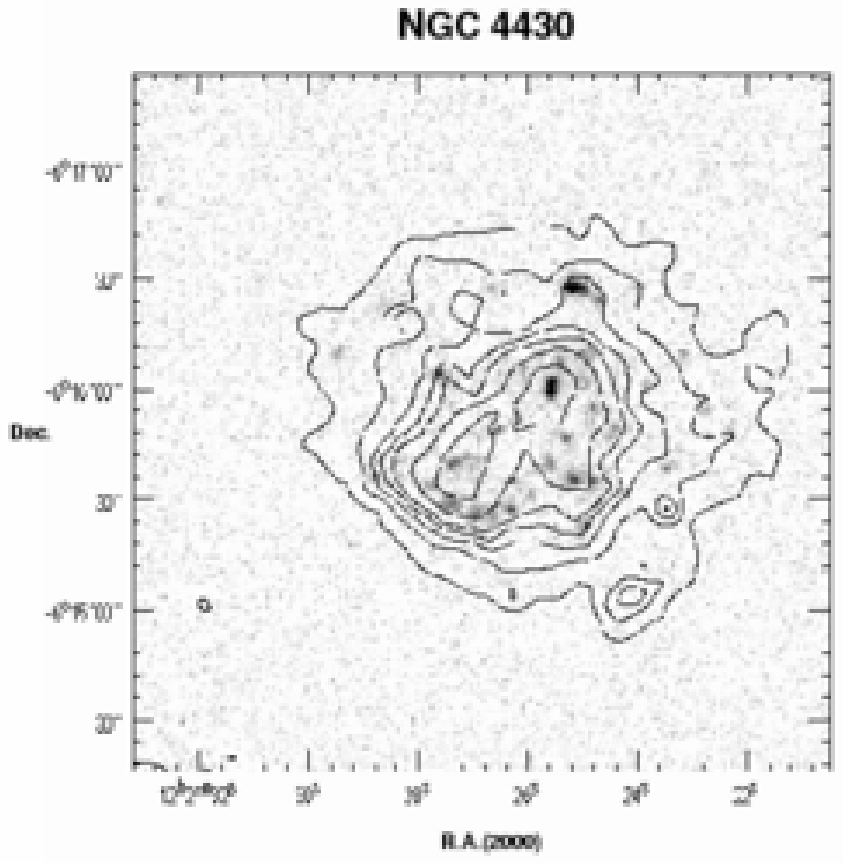}
 \includegraphics[width=7.5cm]{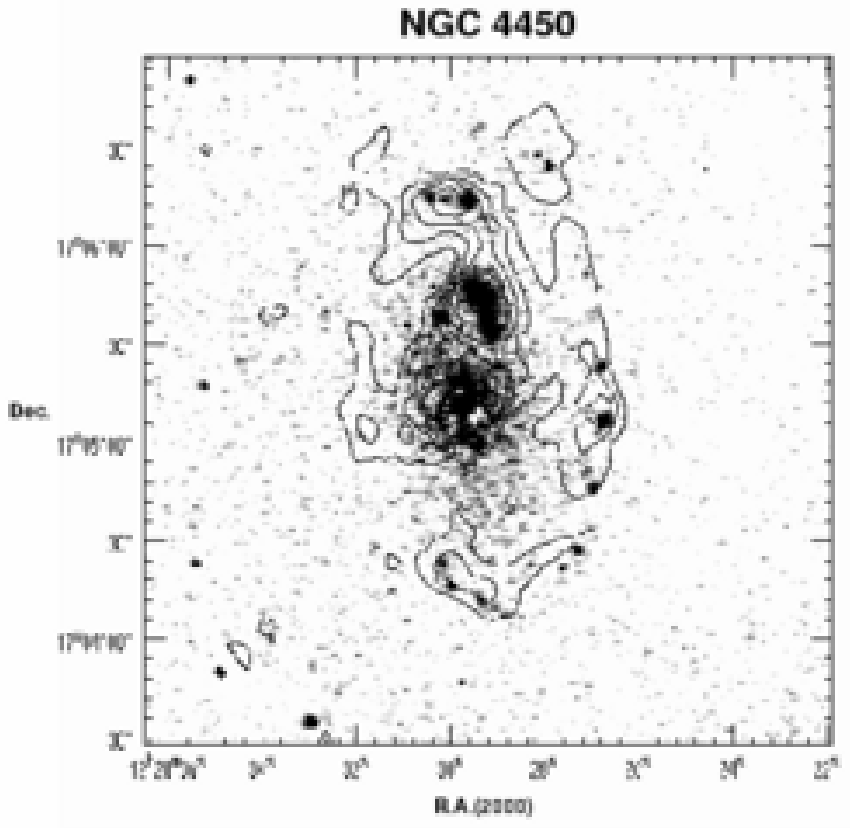}
 \includegraphics[width=7.5cm]{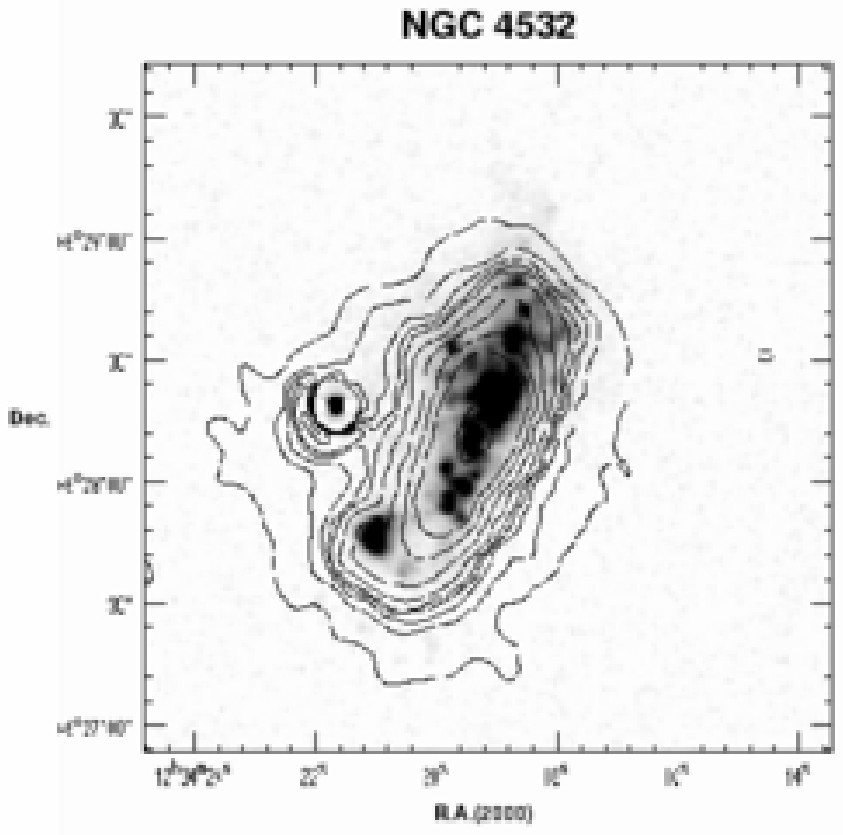}
 \includegraphics[width=7.5cm]{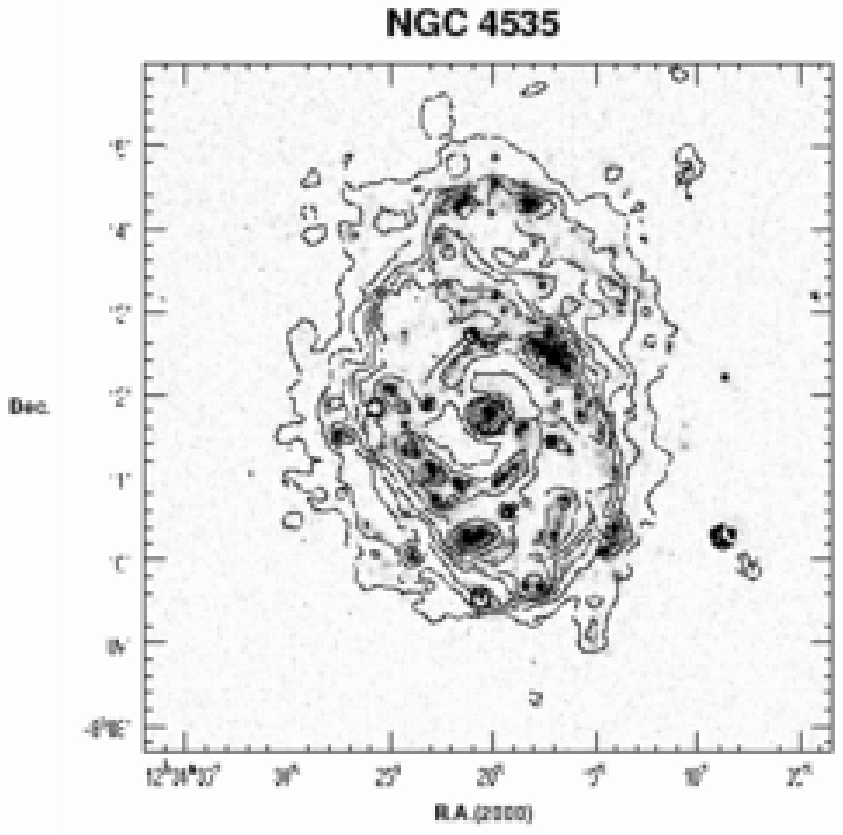}
 \includegraphics[width=7.5cm]{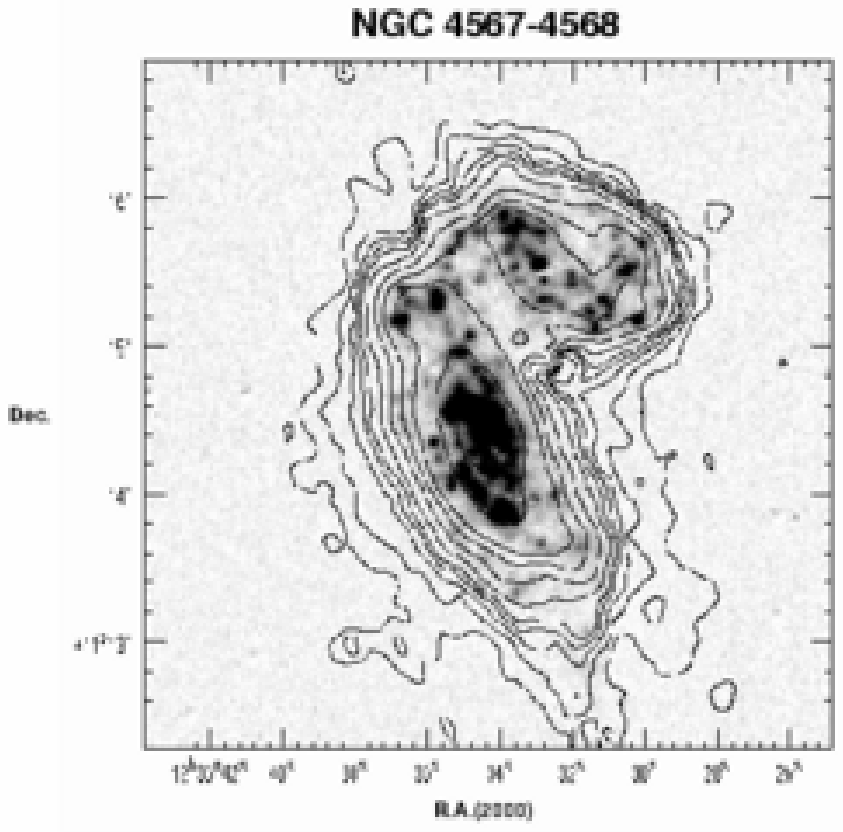}
\caption{The contour plot of the 6.75 $\mu$m dust emission (without the contribution of the stellar
emission) superimposed on the grey scale H$\alpha$+[NII] net image of the resolved galaxies.
Contours are given at 2,4,6,8,10,15,25,40,60,150,300 $\mu$Jy arcsec$^{-2}$. The grey scale is in arbitrary units.} 
 \label{resolved}
 \end{figure*}
 \begin{figure*}[]
 \centering
 \setcounter{figure}{4}
 \includegraphics[width=7.5cm]{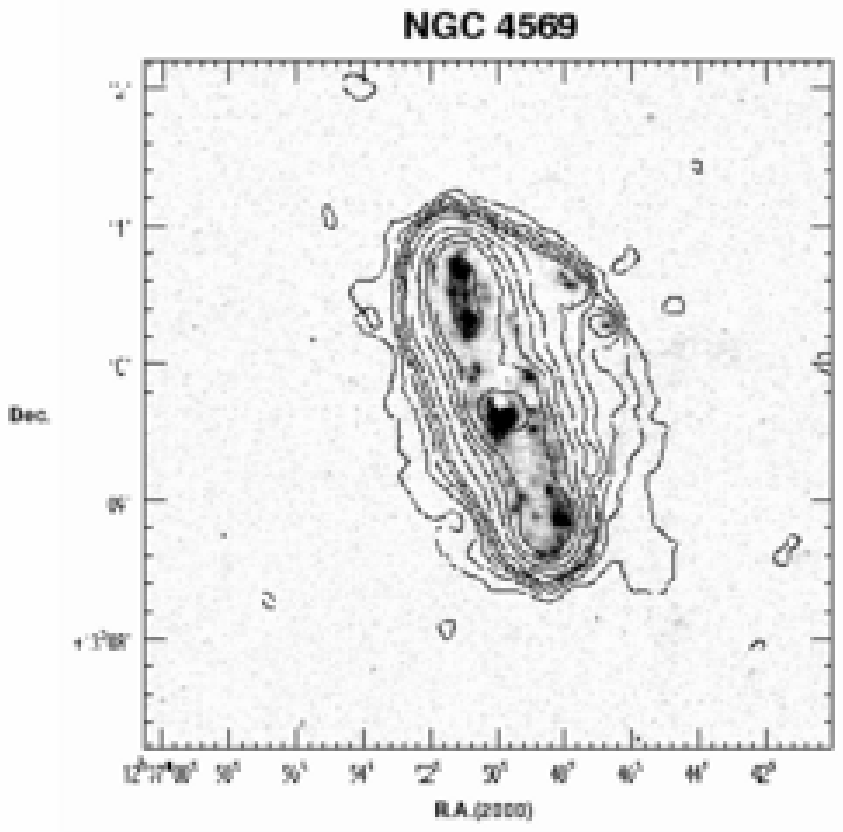}
 \includegraphics[width=7.5cm]{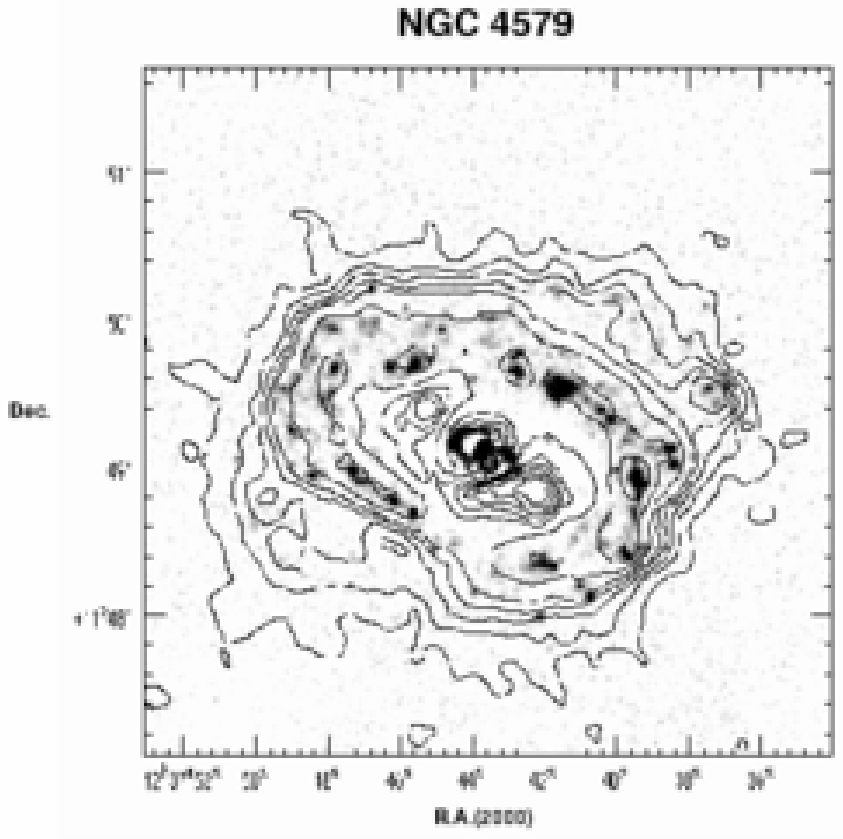}
 \includegraphics[width=7.5cm]{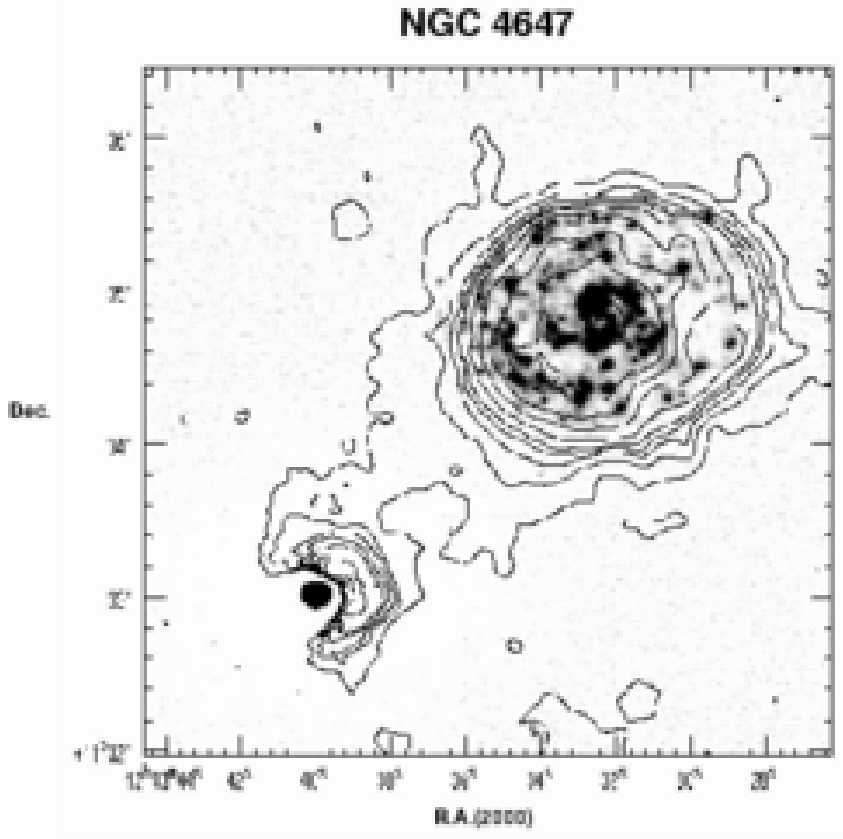}
 \includegraphics[width=7.5cm]{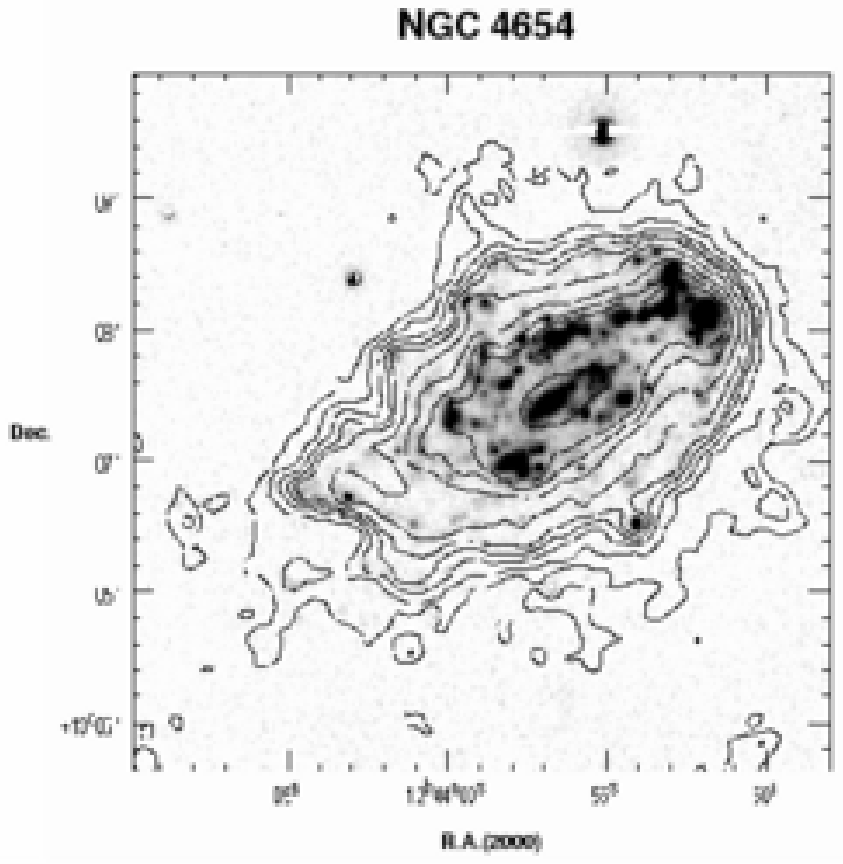}
 \includegraphics[width=7.5cm]{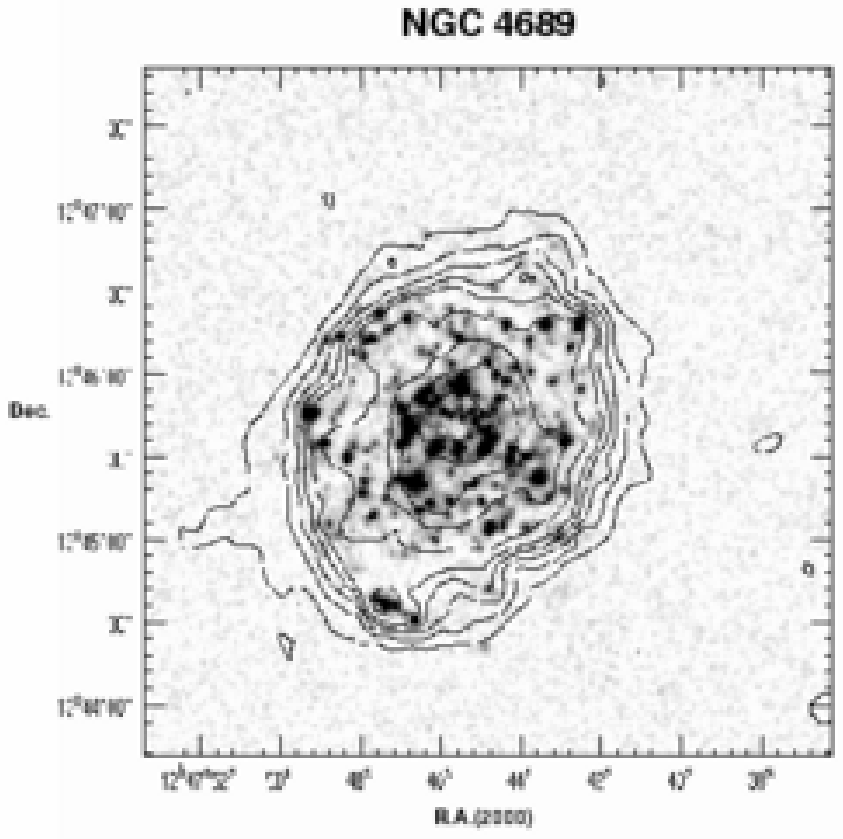}
 \caption{Continue}
 \label{resolved}
 \end{figure*}
\begin{figure*}[]
\centering
\includegraphics[width=12.cm]{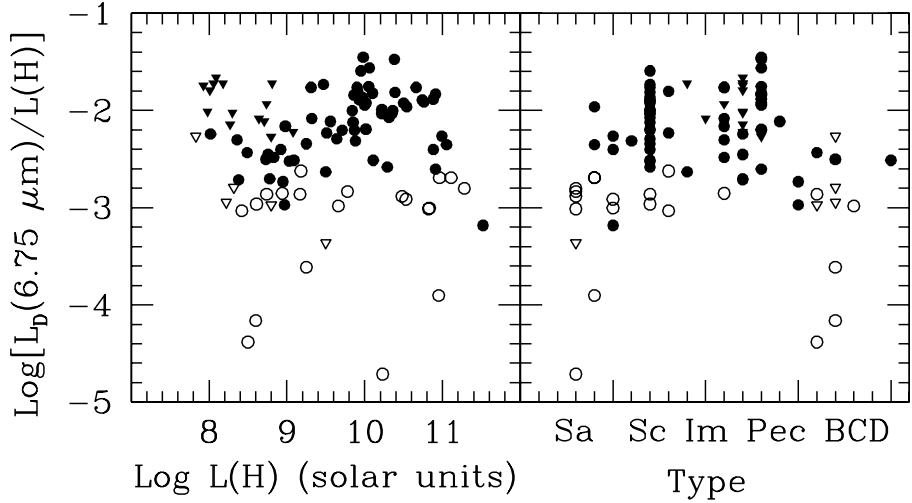}
\caption{The relationship between the normalized (to the H band) 
dust emission mid-IR luminosity at 6.75 $\mu$m and the H band luminosity and the morphological type.
Symbols as in Fig. 2}
\label{camlhty}
\end{figure*}
\noindent

\subsection{Comparison with previous results}

Boselli et al. (1997a; 1998) observed an 
anti-correlation between the mid-IR dust emission and the UV to H band flux ratio in the high star-formation regime.
This behavior is not confirmed by the present analysis, being 
hidden in the high dispersion of the mid-IR vs. H$\alpha$ and vs. UV diagrams of Fig. 2. 
The mid- vs. far-IR relationship just shows a higher dispersion
in the most active galaxies, while the different behavior among star forming and quiescent galaxies
in the mid-IR vs. UV relationships is mostly due to objects whose mid-IR emission
is dominated by stars (empty symbols). All these properties, however, disappear once the pure dust emission is used (Fig. 3).\\
This conclusion seems to be in contrast with the results of Boselli et al. (1998), where mid-IR data
corrected for the stellar contribution were used, weakening (although not excluding) their conclusion
that PAHs are destroyed in strong UV fields. This discrepancy with our 1998 work could be due to: \\
i) the new data reduction pipeline, incorporating a more accurate transient correction 
(see the detailed discussion in Boselli et al. 2003a) gives more reliable mid-IR fluxes
in comparison with the data extracted using the old ISOCAM pipeline that suffer from a systematic overestimate of
the mid-IR fluxes in low luminosity objects.  Low luminosity galaxies are the most active star forming objects
in our sample.\\
ii) UV data in Boselli et al. (1998) were not corrected for dust extinction.\\


The analysis carried out in the previous section has shown that the integrated dust emission of
late-type galaxies in the mid-IR is not fully related to the intensity of the ionizing  
UV radiation field. This result is in agreement with our previous findings based on smaller and less complete samples (Boselli et 
al. 1997a; 1998), but contrasts with the conclusions of Roussel et al. (2001) and F\"orster Schreiber et al.
(2004) who found a tight correlation between the 6.75 and 15 $\mu$m and the H$\alpha$ surface brightness in 
spiral discs. This apparent discrepancy can be due to several factors, that we wish to analyze:\\
1) Systematic sample differences. \\
The sample analyzed by Roussel et al. (2001) includes 49 galaxies, out of which 20 are Virgo members, in common with
our work. Their sample, however, spans the luminosity range -19.2 $\geq$ $M_B$ $\geq$ -21.4, and is thus dominated
by massive galaxies, while our analysis is based on objects of lower average luminosity, with -15 $\geq$ $M_B$ $\geq$ -21.5.
The physical properties of the ISM are expected to change significantly with luminosity: although the 
average UV ionizing field of the two samples is similar (they both span the range -2.3 $\leq$ $log \Sigma H\alpha$ $\leq$ -0.2
L$\odot$ pc$^{-2}$), our sample includes low metallicity objects, with 8.2 $\leq$ $12~+~Log(O/H)$ $\leq$ 9.2.
Metallicity measurements are not available for all the galaxies of Roussel et al. (2001), but given their luminosity
we expect on average solar metallicities (Zaritsky et al. 1994). The sample of F\"orster Schreiber et al. (2004) 
contains the same spirals as Roussel et al. (2001) combined with active starburst and far-IR bright galaxies
with $M_B$ $\leq$ -19.2 and 10$^{42.6}$ $\leq$ $L_{FIR}$ $\leq$ 10$^{45}$ erg s$^{-1}$ and is thus still 
limited to solar metallicity objects. \\
2) Determination of the H$\alpha$ luminosity. \\
H$\alpha$+[NII] imaging data have been corrected for dust extinction 
and [NII] contamination using different recipes. While Roussel et al. (2001) (and thus F\"orster Schreiber et al. (2004)
for the subsample of spiral discs) applied a constant correction both for the [NII] contamination ([NII]/H$\alpha$=0.33)
and for extinction ($A(H\alpha)$=1.1 mag), our spectroscopic integrated data allowed a more accurate galaxy-by-galaxy 
correction. The use of constant corrections might have introduced systematic biases in the data. As shown by 
Gavazzi et al. (2004), the [NII]/H$\alpha$ ratio increases strongly with luminosity, ranging from 
[NII]/H$\alpha$ $\sim$ 0.1 to [NII]/H$\alpha$ $\sim$ 4 when the luminosity increases from $M_B$ = -15
to $M_B$ =-21.5. In a similar way, the Balmer decrement is not constant, but changes significantly from galaxy to galaxy.
Gavazzi et al. (2004) have shown that $A(H\alpha)$ ranges between 0 and $\sim$ 5 mag in late-type galaxies 
with higher extinctions in redder objects. We are thus confident that our H$\alpha$ data are of
higher quality than those for the disc galaxies of Roussel et al. (2001) 
\footnote{For several starburst galaxies the extinction given by F\"orster Schreiber et al. (2004) relies on
near-IR hydrogen recombination lines.}. It is however
unclear why a constant correction of the H$\alpha$ data should reduce the scatter in the mid-IR vs. 
H$\alpha$ relations shown in these two works.\\
3) Normalization.\\ 
To compare galaxies of different size, mid-IR and H$\alpha$ luminosities have been normalized
by the optical area in Roussel et al. (2001) and F\"orster Schreiber et al. (2004) and by the H band luminosity in this work.
We prefer to use the dust free \footnote{In dust-rich objects such as ultra luminous IR galaxies or starburst galaxies 
analysed in F\"orster Schreiber et al. (2004), extinction can be important even in the H band.}
 H band luminosity not only because it is a direct tracer of the total dynamical mass
of normal, late-type galaxies (Gavazzi et al. 1996a), but also because it can be determined with a significantly higher accuracy ($\sim$ 10 \%) 
than the optical area, in particular in objects with an irregular morphology ($\sim$ 30-50 \%). However, as in point 2), we would 
expect that these more accurate corrections and normalizations would lower, instead of increase, the scatter of the 
mid-IR vs. H$\alpha$ relation.\\
4) Correction for stellar contribution.\\ 
Roussel et al. (2001) and F\"orster Schreiber et al. (2004) did not correct the mid-IR luminosities for the stellar contribution. 
While for far-IR and starburst galaxies this
contribution is probably negligible, in quiescent late-type galaxies 
\footnote{Seventeen out of the 49 galaxies in Roussel et al. (2001) (the Virgo galaxies) have a direct measure of the 
stellar contamination in the mid-IR in Boselli et al. (2003b)} the stellar emission can be relevant particularly 
at 6.75 $\mu$m. A higher dispersion in our mid-IR vs. H$\alpha$ relations with respect to those of Roussel et al. (2001)
is however present also when uncorrected data are used (see Fig. 2).\\
5) Disc vs. nucleus. \\
To study the disc properties, Roussel et al. (2001) subtracted the
contribution of the nucleus from both mid-IR and H$\alpha$ data, decreasing the dispersion 
in the mid-IR vs. H$\alpha$ relation (see their Fig. 5). Because of the small angular extent of our sources we cannot
apply a similar aperture correction; the stellar contamination correction that we use, however, should be 
to a first order equivalent since the stellar contribution to the mid-IR emission is dominant in the central part 
of late-type galaxies (Boselli et al. 2003a).\\
In conclusion we believe that the higher observed scatter in our mid-IR vs. H$\alpha$ (normalized) relation with respect to that
observed by Roussel et al. (2001) and F\"orster Schreiber et al. (2004) is real and probably
related to the diversity of the samples. 

\subsection{The origin of the scatter in the mid-IR vs. SFR relations}

To further investigate the origin of the scatter in the mid-IR vs. SFR relationships, 
we look for any systematic trend between 
the residual of the mid-IR pure dust 6.75 $\mu$m vs. H$\alpha$ normalized relationship, 
($Log L_{D 6.75 \mu m}/L_H$=1.105$\times$$Log L_{H\alpha}/L_H$ + 0.504), and any other parameters tracing the physical properties
of the target galaxies. To rule out environmental effects, which in fact are not expected 
since the dust heating in the mid-IR is localized well inside the unperturbed optical disc of the galaxy, we checked that
the residual of the mid-IR vs. H$\alpha$ relationship ($\Delta <L_{D 6.75 \mu m}/L_H - L_{H\alpha}/L_H>$)
is independent of the HI-deficiency parameter \footnote{The HI-deficiency parameter is defined as the ratio of the HI mass to the
average HI mass of isolated objects of similar morphological type and linear size (Haynes \& Giovanelli 1984), which
is a good quantitative indicator of the galaxy-cluster ongoing interaction}.\\
The residual of the mid-IR vs. H$\alpha$ relationship is instead found to depend on the metallicity of the target galaxies, as shown
in Fig. 7.
\begin{figure}[]
\centering
\includegraphics[width=7.cm]{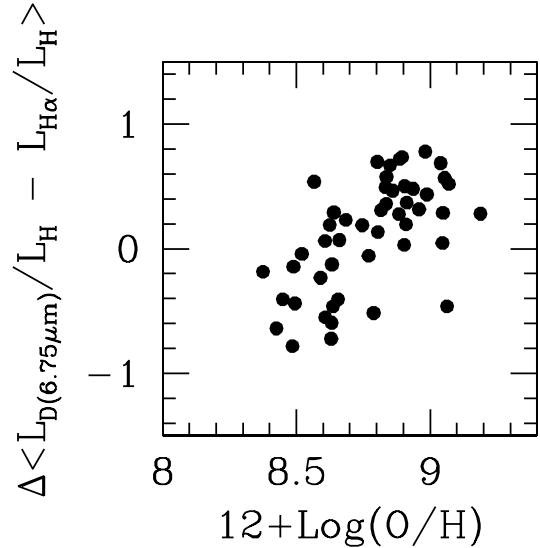}
\caption{The relationship between the residual of the pure dust $L_{D 6.75 \mu m}$ vs. H$\alpha$ normalized luminosities, 
(on logarithmic scales) and the metallicity $12 + Log O/H$ for detected galaxies with a mid-IR dust dominated emission.}
\label{lhauv}
\end{figure}
\noindent
In spite of the large dispersion, Fig. 7 shows that
metal rich galaxies ($12~+~Log(O/H) \sim 9$) have, on average, $\sim$ 10 higher 6.75 $\mu$m mid-IR dust emission per 
unit ionizing UV photons than metal poor objects ($12~+~Log(O/H) \sim 8.4$). The 
higher dispersion in our mid-IR vs. H$\alpha$ relation with respect to that of Roussel et al. (2001) 
and F\"orster Schreiber et al. (2004) can be ascribed to this effect since the mid-IR spectral properties of late-type
galaxies are known to change with metallicity and our sample spans a broader metallicity range (see Fig. 8). 
As summarized by Madden (2000), the intensity
of the UIBs in the 5-20 $\mu$m regime and their relative contribution with respect to the underlying continuum 
\footnote{We recall that, as discussed in F\"orster Schreiber et al. (2004), the emission
of normal galaxies in the 6.75 $\mu$m LW2 ISO filter is generally dominated by the UIBs at 6.2, 7.7 and 8.6 $\mu$m, 
while both UIBs (the strong 12.7 $\mu$m line and minor features at 13.55, 14.25 and 15.7 
$\mu$m, probably dominating in low star formation regimes) and the underlying continuum due to VSGs contribute
to the emission in the 15 $\mu$m LW2 ISO filter.}  
decrease in low metallicity environments, as observed in BCD galaxies (Thuan et al. 1999; Galliano et al. 2003), 
in Wolf-Rayet galaxies 
(Crowther et al. 1999), and in compact HII regions in the Magellanic Clouds (Contursi et al. 2000). In the Small
Magellanic Cloud the UIBs are weak (Li\& Draine 2002b) with the exception of one quiescent cloud (Reach et al. 2000).
Strong UIBs are instead present in metal rich environments with strong UV radiation fields like those 
encountered in luminous starburst galaxies
(M82, NGC 253, NGC 1808: F\"orster Schreiber et al. 2003), although their relative contribution
to the mid-IR emission might decrease in extremely active systems (Lu et al. 2003). 
This can be interpreted by a lower abundance of PAHs due to the lower metallicity, or by their higher destruction
by the UV field which propagates more easily due to the lower dust content, or by differences in the properties
of the UIB carriers (see for the high-metallicity galaxy M31, Cesarsky et al. 1998).
The 6.75 to 15 $\mu$m broad band filter ratio can be used to trace, though only qualitatively, 
the relative importance of the UIB to VSG emission since, as previously discussed, the ISOCAM LW3 15 $\mu$m filter
encompasses both UIB features and dust continuum, while the LW2 6.75 $\mu$m filter 
is generally dominated by UIB emission.\\
Figure 9 does not show any evident trend of the mid-IR flux ratio with the intensity of the UV ionizing field 
(as traced by the H$\alpha$ luminosity) or the metallicity (see also Fig. 1c in Roussel et al. (2001)). 
We should remember, however, that the dynamic ranges of both
the UV radiation field and metallicity sampled with our data are relatively small, preventing the determination of
strong trends. Systematic variations in the UIB to VSG contribution to the mid-IR emission of galaxies because
of depletion of the UIB carriers are in fact expected for radiation fields $\sim$ 10$^{3.5}$ higher than in the solar 
neighborhood (F\"orster Schreiber et al. 2004), observed only in extremely metal poor objects such as IIZw40 
($12~+~Log(O/H)=8.10$; Diaz \& Perez-Montero 2000), NGC 1569 ($12~+~Log(O/H)=8.19$; Kobulnichky \& Skillman 1997)
or SBS0335-052 ($12~+~Log(O/H)=7.33$; Izotov et al. 1997) but not in NGC 1140, where the metallicity is
only $12~+~Log(O/H)=8.46$ (Guseva et al. 2000) (Madden 2000).\\

\subsection{The mid-IR emission as a star formation tracer}

The observational evidence discussed in the previous sections brings us to the conclusion
that, despite different physical properties of the emitting dust, in normal galaxies 
it is roughly the same stellar population that is responsible for the heating of the dust emitting 
both in the mid- and far-IR. Since this population includes stars of different ages and masses,
the first consequence is that, for these quiescent objects, the mid-IR emission is 
not a good tracer of the star formation. However, since the mid-IR emission is often used in the literature
to measure the star formation, we try here to quantify the uncertainty introduced by this method.\\
A calibration based on radiative transfer models at these wavelengths, such as that used by Kennicutt (1998b) 
in the far-IR, is not suited, not only because of the poorly known physical properties of the emitting dust, but also
because dust is stochastically heated and not in thermal equilibrium with the radiation.
We thus derive an empirical calibration by comparing the mid-IR luminosity
to the luminosity in other bands that traditionally are assumed as star formation tracers, as in Boselli et al. (2002b).
Since star formation estimates from mid-IR data generally apply to mid-IR selected samples,
for which accurate stellar SED are unavailable, in this case we use total (uncorrected for 
stellar contamination) mid-IR luminosities.
Fig. 10 shows the relationship between the mid-IR 6.75 and 15 $\mu$m luminosities and the
H$\alpha$, the UV and the far-IR luminosities for galaxies in our sample.
\begin{figure*}[]
\centering
\includegraphics[width=15.cm]{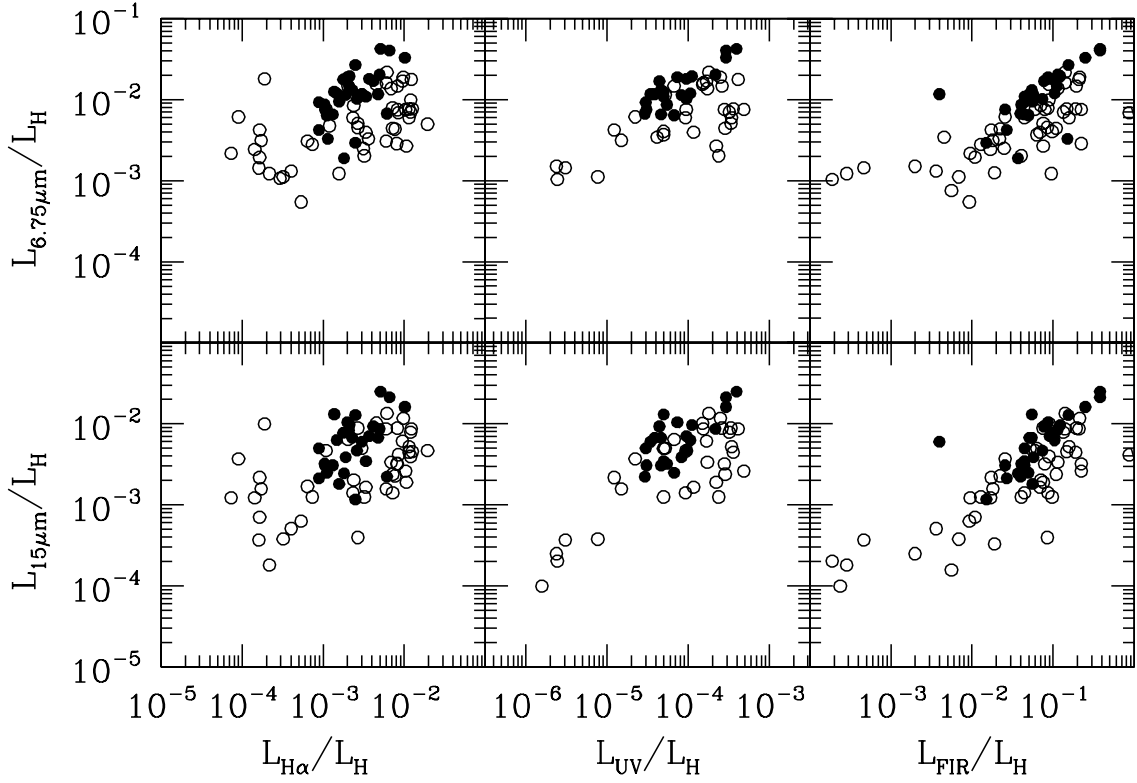}
\caption{The relationship between the normalized (to the H band) mid-IR luminosities at 6.75 (upper) and 
15 $\mu$m (lower) and the H$\alpha$, the UV and the far-IR luminosity for detected galaxies. Filled dots
are for galaxies with solar metallicities ($12 + Log O/H$ $>$ 8.8), empty symbols for sub-solar values ($12 + Log O/H$$\leq$8.8)}
\label{}
\end{figure*}
\noindent
\begin{figure*}[]
\centering
\includegraphics[width=12.cm]{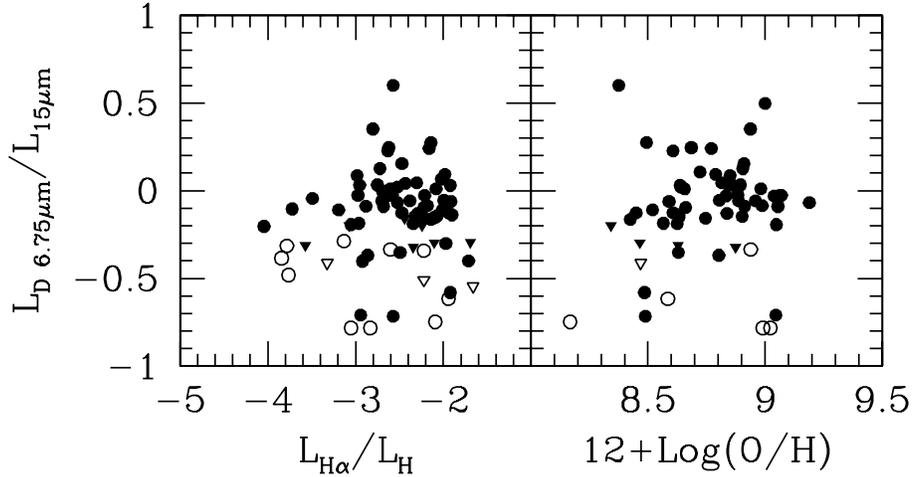}
\caption{The relationship of the mid-IR color 6.75/15 $\mu$m flux ratio with the star formation activity (left)
and with the metallicity (right). Symbols as in Fig. 2.}
\label{}
\end{figure*}
\begin{figure*}[]
\centering
\includegraphics[width=15.cm]{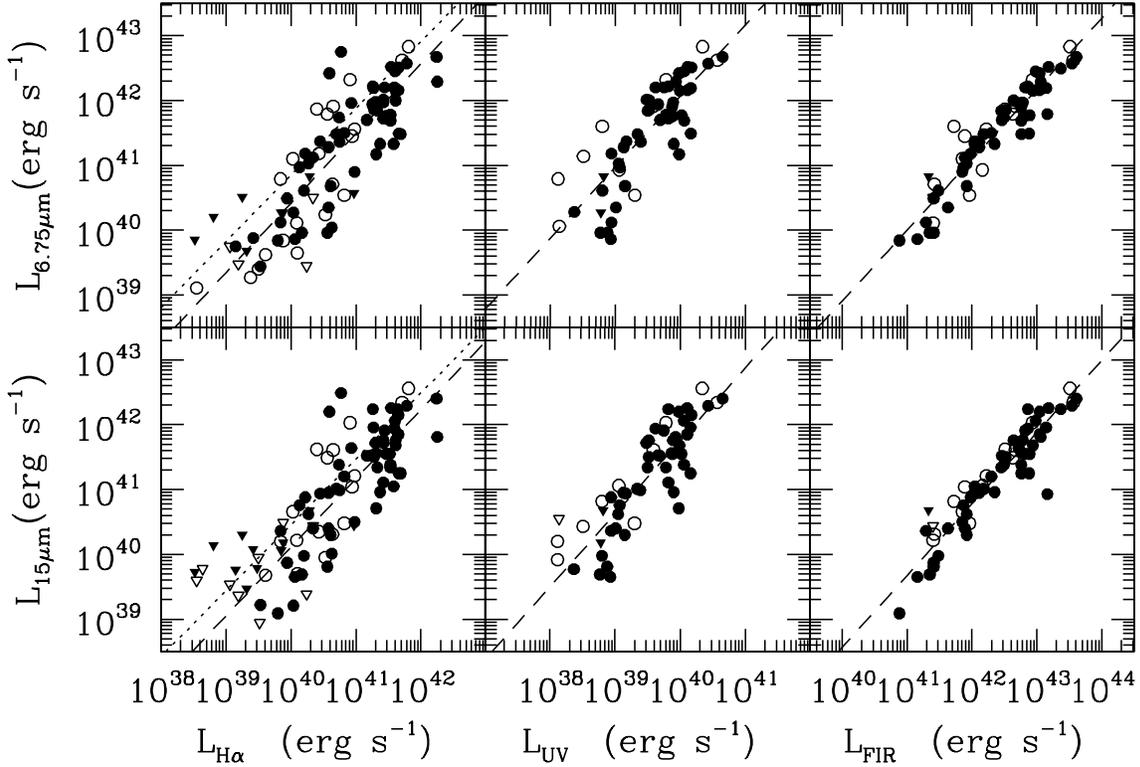}
\caption{The relationship between the mid-IR luminosities at 6.75 (upper) and 
15 $\mu$m (lower) and the H$\alpha$, the UV and the far-IR luminosity.
Filled symbols are for those objects whose mid-IR emission is dominated by dust 
($L_D+L_S/L_S(6.75 \mu m)$ $>$2), open symbols for galaxies whose mid-IR emission 
is mostly stellar ($L_D+L_S/L_S(6.75 \mu m)$ $\leq$ 2).
Triangles indicate upper limits to the mid-IR luminosities. The dashed line indicates the best fit 
given in Table 2, the dotted line the fit given in Roussel et al. (2001) corrected for the $f$ factor (see Sect. 3).}
\label{camsfr}
\end{figure*}
\noindent
As in Fig. 1, the strong relationship is just a scaling effect (none of the variables is 
normalized). 
We want to stress that some of the relationships are non-linear, as can be seen from the lack of a direct
relation with star formation, and that they are all affected by a large dispersion, particularly when the mid-IR luminosity 
is plotted vs. the H$\alpha$ and UV luminosities, confirming the large uncertainty of the
calibration. The best fit to the data, obtained from a bisector linear regression is given in Table 2.\\
\begin{table*}
\caption{Best fits to the relations between the mid-IR 6.75 and 15 $\mu$m luminosities (x variable)
and the three star formation tracers (y variable), with both variables on logarithmic scales.}
\label{Tab2}
\[
\begin{array}{lcrcc}
\hline
\noalign{\smallskip}
{\rm Variable}& {\rm slope} & {\rm constant} & {\rm n.~of~objects} & {\rm R^{2a}}  \\
\noalign{\smallskip}
\hline
\noalign{\smallskip}
           & L(6.75 \mu m)  &                  &   &      \\
\hline
L(H\alpha) & 0.788 \pm 0.040& 8.259 \pm 1.718 & 81& 0.71 \\
L(UV)      & 0.757 \pm 0.051& 8.068 \pm 2.179 & 54& 0.68 \\
L(FIR)     & 0.863 \pm 0.028& 6.568 \pm 1.258 & 62& 0.88 \\
\hline
           & L(15 \mu m)    &                  &   &      \\
\hline
L(H\alpha) & 0.751 \pm 0.049& 10.060 \pm 2.078 & 74& 0.62 \\
L(UV)      & 0.750 \pm 0.045&  8.626 \pm 1.958 & 54& 0.73 \\
L(FIR)     & 0.851 \pm 0.028&  7.322 \pm 1.240 & 62& 0.89 \\
\noalign{\smallskip}
\hline
\end{array}
\]
a: regression coefficient
\end{table*}
\noindent
In the assumption that the star formation activity of these galaxies has been constant
over the last $\sim$ 10$^7$ years for H$\alpha$, $\sim$ 3 10$^8$ years for UV,
the star formation rate of a galaxy (in solar masses per year) can be estimated
from the relationship:
\begin{equation}
{SFR_{\lambda} = K_{\lambda} \times L_{\lambda} ~~~~~~~ {\rm M}_\odot {\rm yr}^{-1}}
\end{equation}
\noindent
Different calibration constants $K_{\lambda}$ valid for  
for several IMFs and metallicities can be found in the literature (Kennicutt 1998b; Boselli et al. 2001;
Iglesias-Paramo et al. 2004), or can be directly estimated from population synthesis models.\\
For example, assuming a Salpeter IMF ($\alpha$=2.35) in the mass range between 0.1 and 100
M$_\odot$, Kennicutt (1998b) gives $K_{{\rm H}\alpha}$ = 1/1.26 10$^{41}$
(M$_\odot$ yr$^{-1}$/erg s$^{-1}$), thus:
\begin{equation}
{SFR = 1.441~ 10^{-33} \times 10^{0.788 \times \log L_{6.75 \mu m}} ~~~~~~~ {\rm
M}_\odot {\rm yr}^{-1}}
\end{equation}
\noindent
We stress that these calibrations apply to late-type 
galaxies in the mid-IR luminosity range between
10$^{39}$ $\leq$ $L(6.75 \mu m)$ $\leq$ 10$^{43}$ erg s$^{-1}$ and  
10$^{39}$ $\leq$ $L(15 \mu m)$ $\leq$ 10$^{42.6}$ erg s$^{-1}$. We do not know
whether these values apply outside this range, in particular
at higher luminosities such as those encountered in far-IR bright galaxies.
We also recall that, as extensively discussed in Kennicutt (1998b), far-IR
luminosities can be transformed into SFR (in M$\odot$ yr$^{-1}$)
only in dusty starburst galaxies. Galaxies analyzed in this work have far-IR
luminosities in the range 10$^{41}$ $\leq$ $L(FIR)$ $\leq$ 10$^{43.7}$ erg s$^{-1}$
(10$^{7.4}$ $\leq$ $L(FIR)$ $\leq$ 10$^{10.1}$ in solar units) and are normal, quiescent late-type galaxies. The use of
the mid- vs. far-IR empirical calibration given in Table 2 would thus lead to
highly uncertain results. 
The dispersion in the $L_{{\rm H}\alpha}$ vs. $L_{6.75 \mu m}$ and $L_{15 \mu m}$ relations 
is a factor of $\sim$ 5 (1 $\sigma$).
The uncertainty in the determination of the SFR of galaxies using Eq.
(4) is still larger. As shown in Charlot \& Longhetti (2001), the
uncertainty in the determination of SFR from H$\alpha$ data using stellar
population synthesis models is already a factor of $\sim$ 3 when the
data are properly corrected for dust extinction and [NII] contamination, as done in this work. 
If we take into account all the possible sources of error in the determination
of the H$\alpha$ luminosity, and the uncertainty introduced
by the large dispersion observed in Fig. 10, we conclude that the
resulting uncertainty in the determination of the SFR from mid-IR luminosity
measurement is as high as a factor of $\sim$ 10. \\

\section{Conclusion}

Using a sample of 123 normal, late-type, nearby galaxies with available multifrequency data we have
studied the relationship between the mid-IR (5-18 $\mu$m) emission and various other star formation tracers 
for investigating the nature of the dust heating sources in this spectral domain. 
This analysis has shown that the normalized mid-IR luminosity correlates better with the far-IR luminosity 
than with more direct tracers of the young stellar population such as the H$\alpha$ and the UV luminosity. 
The comparison of resolved images reveals a remarkable similarity in the H$\alpha$ and mid-IR morphology, with all HII 
regions prominent at both frequencies. However, once corrected for the stellar contamination, mid-IR images are also 
characterized by a diffuse emission not associated with HII regions nor with the diffuse H$\alpha$ emission. \\
Although radiation transfer processes cannot be excluded, this evidence, in agreement with similar results 
obtained for the ISM of our own Galaxy,
indicates that the stellar population responsible for the heating of dust emitting in the mid-IR is similar to that
which heats big grains emitting in the far-IR. The dust emission of a whole galaxy in the 5-18 $\mu$m domain is thus due to dust
heated by both the ionizing and non-ionizing radiation, the latter produced mainly by relatively evolved stars.
The direct consequence of our analysis is that the mid-IR luminosity is not an optimal  
star formation tracer in normal, late-type galaxies.
We further analyzed the origin of the scatter in the mid-IR vs. H$\alpha$, UV and far-IR luminosity relationships
and demonstrated that mostly has to do with metallicity effects, with metal-poor 
objects having a lower mid-IR emission per unit star formation rate than metal-rich galaxies.\\
The present results hold for normal, late-type galaxies and do not necessarely
apply to luminous starburst galaxies or ultra-luminous IR galaxies. These extreme objects are not
very frequent in the nearby Universe but are probably representative of forming 
galaxies detected in mid-IR deep surveys.
F\"orster Schreiber et al. (2004) claim that for actively-star-forming galaxies
the mid-IR luminosity is a good star formation tracer. 
This apparent discrepancy will be resolved once extensive spectro-photometric surveys 
of galactic and extragalactic sources in a variety of physical conditions will 
help to understand the connection between 
the mid-IR dust emission properties (namely the relative contribution of the UIB line and
continuum emission, the destruction of the UIB carriers, the contribution of the very small grains, etc.)
and the physical conditions of the ISM, namely the metallicity and the UV radiation field.

\begin{acknowledgements}      

We would like to thank V. Buat, A. Contursi and J.M. Deharveng for interesting discussions, and 
L. Cortese for the determination of the best fits. We would like to thank the anonymous referee for his comments
and suggestions which helped to improve the quality of the manuscript.

\end{acknowledgements}

\end{document}